# Title: Three-dimensional crystals of adaptive knots

**Authors:** Jung-Shen B. Tai[1], Ivan I. Smalyukh[1,2,3,*]

**Affiliations:**

[1]Department of Physics, University of Colorado, Boulder, Colorado 80309, USA

[2]Materials Science and Engineering Program, Soft Materials Research Center and Department of Electrical, Computer and Energy Engineering, University of Colorado, Boulder, Colorado 80309, USA

[3]Renewable and Sustainable Energy Institute, National Renewable Energy Laboratory and University of Colorado, Boulder, Colorado 80309, USA

*Correspondence to: ivan.smalyukh@colorado.edu

**Abstract:** Starting from Gauss and Kelvin, knots in fields were postulated behaving like particles, but experimentally they were found only as transient features or required complex boundary conditions to exist and couldn't self-assemble into three-dimensional crystals. We introduce energetically stable micrometer-sized knots in helical fields of chiral liquid crystals. While spatially localized and freely diffusing in all directions, they resemble colloidal particles and atoms, self-assembling into crystalline lattices with open and closed structures. These knots are robust and topologically distinct from the host medium, though they can be morphed and reconfigured by weak stimuli under conditions like in displays. A combination of energy-minimizing numerical modeling and optical imaging uncovers the internal structure and topology of individual helical field knots and various hierarchical crystalline organizations they form.

**One Sentence Summary:** Stable solitonic and vortex knots in molecular alignment fields behave like particles and form triclinic crystals.

**Main Text:**

Topological order and phases represent an exciting frontier of modern research (*1*), but topology-related ideas have a long history in physics (*2*). Gauss postulated that knots in fields could behave like particles whereas Kelvin, Tait and Maxwell believed that the matter, including crystals, could be made of real-space free-standing knots of vortices (*2-4*). These early physics models, introduced long before even the very existence of atoms was widely accepted, gave origins to modern mathematical knot theory (*2-4*). Expanding this topological paradigm, Skyrme and others modeled subatomic particles with different baryon numbers as nonsingular topological solitons and their clusters (*3-5*). Knotted fields emerged in classical and quantum field theories (*3-7*) and in scientific branches ranging from fluid mechanics to particle physics and cosmology (*2-11*). In condensed matter, arrays of singular vortex lines and low-dimensional analogs of Skyrme solitons were found as topologically nontrivial building blocks of exotic thermodynamic phases in superconductors, magnets and liquid crystals (LCs) (*12-14*). Could they be knotted, and could these knots self-organize into three-dimensional (3D) crystals? Knotted fields in condensed matter found many experimental and theoretical embodiments, including both nonsingular solitons and knotted vortices (*7-9, 15-23*). However, they were metastable and decayed with time (*7-9,15-17*) or could not be stabilized without colloidal inclusions (*18,19*), confinement and boundary conditions (*20-22*), as well as could not self-organize into 3D lattices (*22,23*). We introduce energetically stable micrometer-sized adaptive knots in chiral LCs that, unexpectedly, materialize the knotted vortices and nonsingular solitonic knots at the same time and indeed behave like particles, undergoing 3D Brownian motion and self-assembling into 3D crystals.

Helical fields, as in the familiar example of circularly-polarized light with electric and magnetic fields periodically rotating around the Poynting vector, are ubiquitous in chiral materials like magnets and LCs. These helical fields comprise a triad of orthonormal fields (Fig. 1A), namely the material alignment field $\mathbf{n}(\mathbf{r})$ (of rod-like molecules in LCs or spins in magnets) and the immaterial line fields along the helical axis $\boldsymbol{\chi}(\mathbf{r})$, analogous to the Poynting vector, and $\boldsymbol{\tau}(\mathbf{r}) \perp \mathbf{n}(\mathbf{r}) \perp \boldsymbol{\chi}(\mathbf{r})$. For LCs, $\mathbf{n}(\mathbf{r})$ is nonpolar but can be decorated with unit vector fields (*14,24*). The distance over which $\mathbf{n}(\mathbf{r})$ and $\boldsymbol{\tau}(\mathbf{r})$ rotate around $\boldsymbol{\chi}(\mathbf{r})$ by $2\pi$ within the helical structure is the helical pitch $p$ (Fig. 1A). We demonstrate knotted fields that in $\mathbf{n}(\mathbf{r})$ are topological solitons with inter-linked closed-loop preimages resembling Hopf fibration (Fig. 1B). At the same time, the nonpolar nature of $\boldsymbol{\chi}(\mathbf{r})$ and $\boldsymbol{\tau}(\mathbf{r})$ permits the half-integer singular vortex lines forming various torus knots (Fig. 1C) while retaining fully nonsingular nature of $\mathbf{n}(\mathbf{r})$. Therefore, our topological soliton in the helical field is a hybrid embodiment of both interlinked preimages and knotted vortex lines, which can be realized to have this solitonic nonsingular nature in systems with either polar or non-polar $\mathbf{n}(\mathbf{r})$ (*12-14*). We find these knot solitons, which we call "heliknotons", embedded in a helical background and forming spontaneously after transition from the isotropic to LC phase when a weak electric field $\mathbf{E}$ is applied to a positive-dielectric-anisotropy chiral LC along the far-field helical axis $\boldsymbol{\chi}_0$. The materials utilized are prepared as mixtures LC-1 through LC-3 (*24*) of commercially available room-temperature nematics and chiral dopants. In bulk LC samples of typical thickness within $d$=10-100μm (*24*), heliknotons display 3D particle-like properties and form a dilute gas at low number densities (Fig. 1D), with orientations of shape-anisotropic solitonic structures correlated with their positions along $\boldsymbol{\chi}_0$ (Fig. 1, D and E). Depending on materials and applied voltage $U$, heliknotons can adopt different shapes (Fig. 1, D to G), which are reproduced by numerical modeling (insets of Fig. 1, F and G) based on minimization of the free energy (*24*):

$$F = \int d^3\mathbf{r} \left\{ \frac{K}{2} (\nabla \mathbf{n})^2 + \frac{2\pi K}{p} \mathbf{n} \cdot (\nabla \times \mathbf{n}) - \frac{\varepsilon_0 \Delta \varepsilon}{2} (\mathbf{n} \cdot \mathbf{E})^2 \right\} \quad (1)$$

where $K$ is the average elastic constant, $\Delta\varepsilon$ is the LC's dielectric anisotropy and $\varepsilon_0$ is the vacuum permittivity. The integrand comprises energy terms originating from elastic deformation, chirality and dielectric coupling, respectively. Minimization of $F$ at different $U$ and $\Delta\varepsilon$ (table S1) reveals that heliknotons can be stable, metastable or unstable with respect to the helical background (Fig. 1H), comprising localized regions (depicted in gray in Fig. 1, B and C) of perturbed helical fields and twisting rate.

Heliknotons undergo Brownian motions (Fig. 1I and movie S1) and exhibit anisotropic interactions while moving along $\boldsymbol{\chi}_0$ and in the lateral directions (Fig. 1, E and J, and movie S2) (*24*). The inter-heliknoton pair interaction potential is anisotropic and highly tunable, from attractive to repulsive and from tens to thousands $k_\mathrm{B}T$, depending on the choice of LC, $U$ and sample thickness (Fig. 1J). Similar to nematic colloids (*25,26*), interactions between localized heliknotons arise from sharing long-range perturbations of the helical fields around them and minimizing the overall free energy for different relative spatial positions of these solitons. These interactions lead to a plethora of crystals, including low-symmetry and open lattices that were recently achieved in colloids (*26-28*) (Fig. 2). In thin cells of thickness $d \lesssim 4p$, heliknotons localize around the sample's horizontal midplane, making their anisotropic interactions quasi-2D. Heliknotons self-assemble (movie S3) into a 2D rhombic lattice both when the attractive potential is ~1000 $k_BT$ (Fig. 2, A and B) and ~10 $k_BT$ (Fig. 2, C and D). From initial positions defined by laser tweezers (*24*), heliknotons self-assemble into a stretched kagome lattice with anisotropic binding energies ~100 $k_BT$ (Fig. 2E). Such open lattices have interesting topological properties (*28*), potentially bringing about an interplay between topologies of the crystal's basis and lattice. Crystallographic symmetries and lattice parameters can be controlled through tuning reconfigurable interactions, like switching reversibly between synclinic and anticlinic tilting of heliknotons via changing $U$ by <0.5V (Fig. 2, F and G, movie S4).

3D crystals of heliknotons emerge at $d > 4p$, when anisotropic interactions yield triclinic lattices (Fig. 2, H to N). One can watch initially quasi-2D pre-self-assembled crystallites interacting with each other while moving in lateral and axial directions (Fig. 2, J to M, and movie S5), forming different crystallographic planes of the 3D triclinic lattice. The helical background LC, individual heliknotons and the ensuing lattices are all chiral. The lowest-symmetry triclinic pedial lattices can have primitive cells comprising two (Fig. 2, H and J to M, and movie S5) or three crystallographic planes (Fig. 2, I and N, and movie S6), depending on relative orientations of heliknotons within these planes. The two lattices with parallel (Fig. 2M) and orthogonal (Fig. 2N) relative orientations of heliknotons in consecutive heliknoton layers are just examples as the angle between heliknotons within crystallographic planes along $\chi_0$ can be tuned (Fig. 1E) by $U$, material and geometric parameters. Since the heliknotons have anisometric shape (like LC molecules) and can exhibit spatial twists, potentially even hierarchical topological solitons comprising heliknotons could emerge. Heliknoton crystals exhibit giant anisotropic electrostriction (Fig. 2O and movies S7 and S8). For example, upon changing from $U$=3.0 to 4.4V, one lattice parameter in the insets of Fig. 2O extends by ~44% while the other only by ~4%. This electrostriction is consistent with the free energy minimization (Fig. 2P) for 3D crystals of heliknotons with tunable lattice parameters at different $U$. The experimentally observed soliton crystals correspond to minima of free energy within a broad range of applied voltages (*24*), consistent with their facile self-organization into triclinic pedial crystals and other reconfigurable 3D and 2D lattices (Fig. 2). As the applied field is increased even further, the heliknoton crystals become metastable and then unstable with respect to the unwound state with $\mathbf{n}(\mathbf{r}) \| \mathbf{E}$ (*24*).

Numerical modeling and experiments reveal detailed structures of the fields within a heliknoton (*24*) (Fig. 3, A to I, and fig. S1). In $\mathbf{n}(\mathbf{r})$, the continuous localized configuration of a heliknoton is embedded in a helical background (Fig. 3, A to C) and has all closed-loop preimages linked with each other once, with the linking number +1 (Fig. 3J and fig. S1). Up to numerical precision, this matches the Hopf index calculated through numerical integration (*22*),

$$Q = \frac{1}{64\pi^2} \int d^3\mathbf{r} \, \epsilon^{ijk} A_i B_{jk}, \quad (2)$$

where $B_{ij} = \epsilon_{abc} n^a \partial_i n^b \partial_j n^c$, $\epsilon$ is the totally antisymmetric tensor, $A_i$ is defined as $B_{ij} = (\partial_i A_j - \partial_j A_i)/2$, and the summation convention is assumed. Spatial structures of $\chi(\mathbf{r})$ and $\tau(\mathbf{r})$ are derived from the energy-minimizing $\mathbf{n}(\mathbf{r})$ using the eigenvector of the chirality tensor (*24,29,30*) (Fig. 3, D to I). They exhibit torus knots of spatially co-located singular half-integer vortices, within which $\chi(\mathbf{r})$ and $\tau(\mathbf{r})$ nonpolar fields rotate by 180° around the vortex line in the plane locally orthogonal to it (Fig. 3, D to I). The closed loop of the vortex line is the righthanded T(2,3) trefoil torus knot, also labeled as the $3_1$ knot in the Alexander–Briggs notation (Fig. 3, K and L). The singular vortex knots in $\chi(r)$ and $\tau(r)$ also correspond to a co-located knot of a meron (topologically nontrivial structure of a fractional 2D skyrmion tube) in $n(r)$ (fig. S2). Handedness of the knots and links matches that of chiral $n(r)$, implying that the sign of Hopf indices of such energy-minimizing solitons is dictated by LC's chirality. Simulated and experimental depth-resolved nonlinear optical images of heliknotons for different polarizations of excitation light closely agree (Fig. 3, M to O), confirming experimental reconstruction of the field (*24*). Unlike the Shankar solitons (*11*), which exemplify condensed matter models with topology of a triad of orthonormal fields similar to that of Skyrme solitons in nuclear physics, heliknotons exhibit nonsingular structure only in one of the three fields, though they are still overall nonsingular in the material field. Differing from transient textures of linked loops of nonsingular disclinations (*15,16*) and metastable loops of singular vortices (*17*) in cholesteric LCs, our heliknotons are stable torus knots of co-located merons in $n(r)$ and vortices in $\chi(r)$ and $\tau(r)$ that enable ground-state 3D crystals of knots (Fig. 2 and fig S2).

Besides the $Q = 1$ heliknotons with equilibrium dimensions between $p$ and $2p$, we also find larger $Q = 2$ topological solitons (Fig. 4A), for which experimental polarizing micrographs also match their numerical counterparts. A $Q = 2$ heliknoton contains a larger region of distorted helical background in both lateral and axial directions (Fig. 4, A to D, figs. S3 and S4). Preimages for two anti-parallel vertical orientations of $n(r)$ form a pair of Hopf links (Fig. 4H), linked twice, like all other preimage pairs. Singular vortex lines in $\chi(r)$ and $\tau(r)$ form closed cinquefoil T(2,5) torus knots (also labeled as $5_1$ knots) co-located with a similar knot of a meron tube in $n(r)$. A $Q = 3$ heliknoton contains three Hopf links of preimages with a net linking number of 3 for each preimage pair (Fig. 4I, and figs. S5 and S6). The singular vortex lines in $\chi(r)$ and $\tau(r)$ form a T(2,7) torus knot (the $7_1$ knot), co-located with the same knot of a meron tube in $n(r)$ (Fig. 4, E, F and I). Figure 4, G to I show both preimages of the anti-parallel vertical orientations of $n(r)$ and vortex lines in $\chi(r)$ and $\tau(r)$, as well as the Reidemeister moves simplifying their structures. Interestingly, topologically distinct heliknotons have different numbers of crossings in the free-standing knots of vortex lines and different linking of preimages, which were key topological invariants in early models of atoms and subatomic particles (*2-6*). For different heliknotons, $Q$ is related to crossing number $N$ of the vortex knots: $N = 2Q + 1$. Remarkably, the closed loop of preimages in $n(r)$ are inter-linked with the torus knots of vortices in $\chi(r)$ and $\tau(r)$, as shown in Fig. 4, G to I.

Differing from transient vortex lines, which shrink with time due to energetically costly cores and distorted order around them, vortex-meron knots in heliknotons are energetically favorable because of being nonsingular in the material $n(r)$ field and comprising twisted structures with handedness matching that of the LC. The stability of our 3D topological solitons as spatially localized structures is assisted by the chiral term in Eq. (1), which introduces their finite dimensions and plays a role analogous to that of high-order nonlinear terms in solitonic models of subatomic particles (*3-6*) and the Dzyaloshinskii-Moriya term in models of magnetic skyrmions (*13,14*). Applied field along $\chi_0$ tends to reorient $n(r)$ along $E$ as compared to the helical state with $n(r) \perp E$ and knots emerge as local or global energy minima within a certain range of voltages (*24*) by reducing the dielectric term in Eq. (1) as compared to the helical state (Figs. 1H and 2P). At low $E$, elastic energetic costs are high and heliknotons collapse into the helical background through spontaneous creation and annihilation of singular defects in $n(r)$.

The strong dielectric coupling between $\mathbf{n}(\mathbf{r})$ and $\mathbf{E}$ aligns $\mathbf{n}(\mathbf{r})\|\mathbf{E}$ at high applied fields, eventually making both the helical structure and solitons unstable, but heliknotons are the global free energy minima within broad material-dependent ranges of $\mathbf{E}$ (*24*).

We have demonstrated 3D topological solitons in helical fields of chiral LCs that can be ground-state and metastable configurations, forming 3D crystalline lattices. Unlike the atomic, molecular and colloidal crystals, heliknoton crystals exhibit giant electrostriction and dramatic symmetry transformations under <1V voltage changes. We envisage that such solitons can emerge in helical phases of solid-state non-centrosymmetric magnets (*13,14,22*) and ferromagnetic LCs (*23*) with helical fields and Hamiltonians similar to those of chiral LCs, where the roles of dielectric and chiral terms in Eq. (1) can be played by magnetocrystalline anisotropy and Dzyaloshinskii-Moriya interactions (*23*), respectively. Our crystals are experimental embodiments of matter made of Kelvin's vortex knots and Skyrme's knot solitons.

**Acknowledgments:** We thank P. Ackerman, M. Dennis, T. Lubensky, N. Manton, B. Senyuk, Y. Shnir, H. Sohn, R. Voinescu and Y. Yuan for discussions.

**Funding:** This work was supported by the National Science Foundation grant DMR-1810513.

**Author contributions:** J.-S.B.T. and I.I.S. performed research and wrote the manuscript. I.I.S. conceived the project and provided funding.

**Competing interests:** The authors declare no competing interests.

**Data and materials availability:** All data are reported in the main text and supplementary materials.

**Supplementary Materials:**

Materials and Methods

Figures S1-S6

Table S1

Movies S1-S8

An MS Excel file with the data used to obtain plots in figures

References (*31-47*)

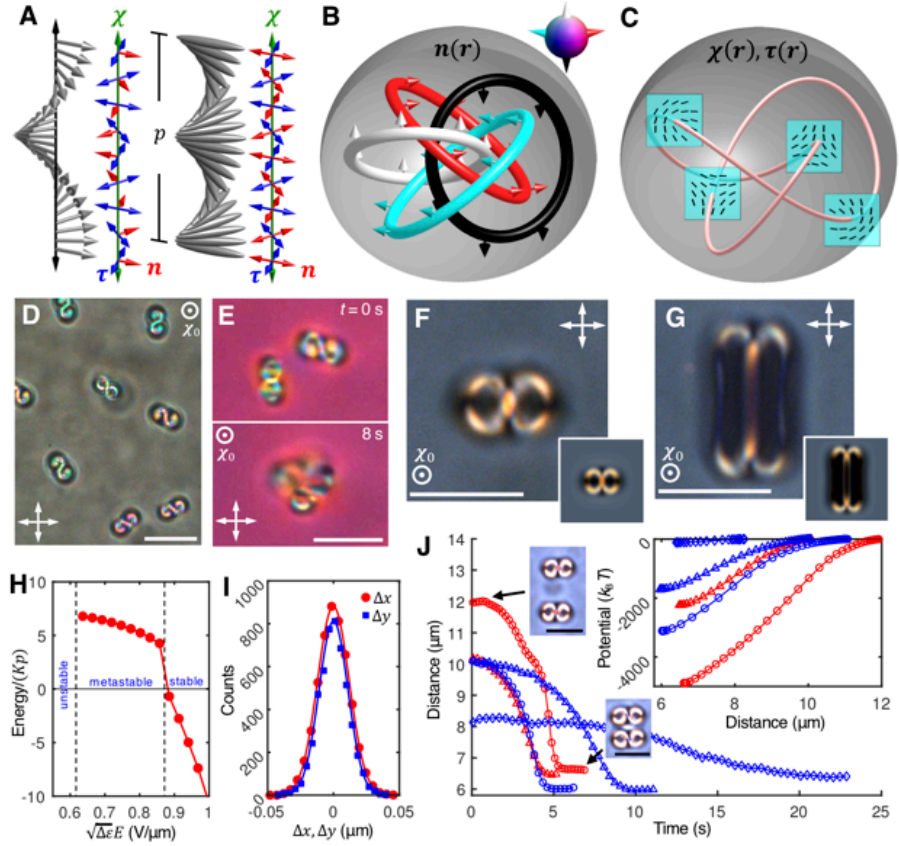

**Fig. 1. Knots in helices.** (**A**) Helical field comprising a triad of orthonormal fields $\boldsymbol{n(r)}$, $\boldsymbol{\chi(r)}$, $\boldsymbol{\tau(r)}$ with $\boldsymbol{n(r)}$ being either polar (left) or nonpolar (right). (**B**) Preimages in $\boldsymbol{n(r)}$ of a heliknoton colored according to their orientations on $\mathbb{S}^2$ (top-right inset). (**C**) Knotted co-located half-integer vortex lines in $\boldsymbol{\chi(r)}$ and $\boldsymbol{\tau(r)}$. In (B) and (C), the gray isosurfaces show regions of distorted helical background. (**D**) A gas of heliknotons in LC-1 sample of thickness 30 µm at $U$=4.5V. (**E**) Two heliknotons interact in 3D while forming a dimer in LC-2 sample of thickness 30 µm at $U$=11.0V. (**F** and **G**) Polarizing optical micrographs of metastable and stable heliknotons at $U$=4.3V and 4.5V, respectively, in a sample with $d = 10$µm, with computer-simulated counterparts shown in the bottom-right insets. (**H**) Free energy of individual heliknotons versus $\boldsymbol{E}$, where energy of the helical state equals zero; the helical state and heliknotons are unstable at $\sqrt{\Delta\varepsilon}E \gtrsim 1$V/µm when the field tends to align $\boldsymbol{n(r)}\|\boldsymbol{E}$ (*24*). (**I**) Displacement histograms Δx and Δy showing diffusion of the heliknoton in (F) in orthogonal lateral directions perpendicular to $\boldsymbol{\chi_0}$. Experimental and numerical data were obtained for LC-1 in (D) and LC-2 in (E) to (I). (**J**) Pair interaction of heliknotons. Data shown in red (at voltages ○: 4.2V, △: 2.9V) were obtained for LC-2 and in blue (○:1.4V, △:1.0V, ◇:1.7V) for LC-1 at $d≈10$µm (*24*). Scale bars are 10µm and $p = 5$µm.

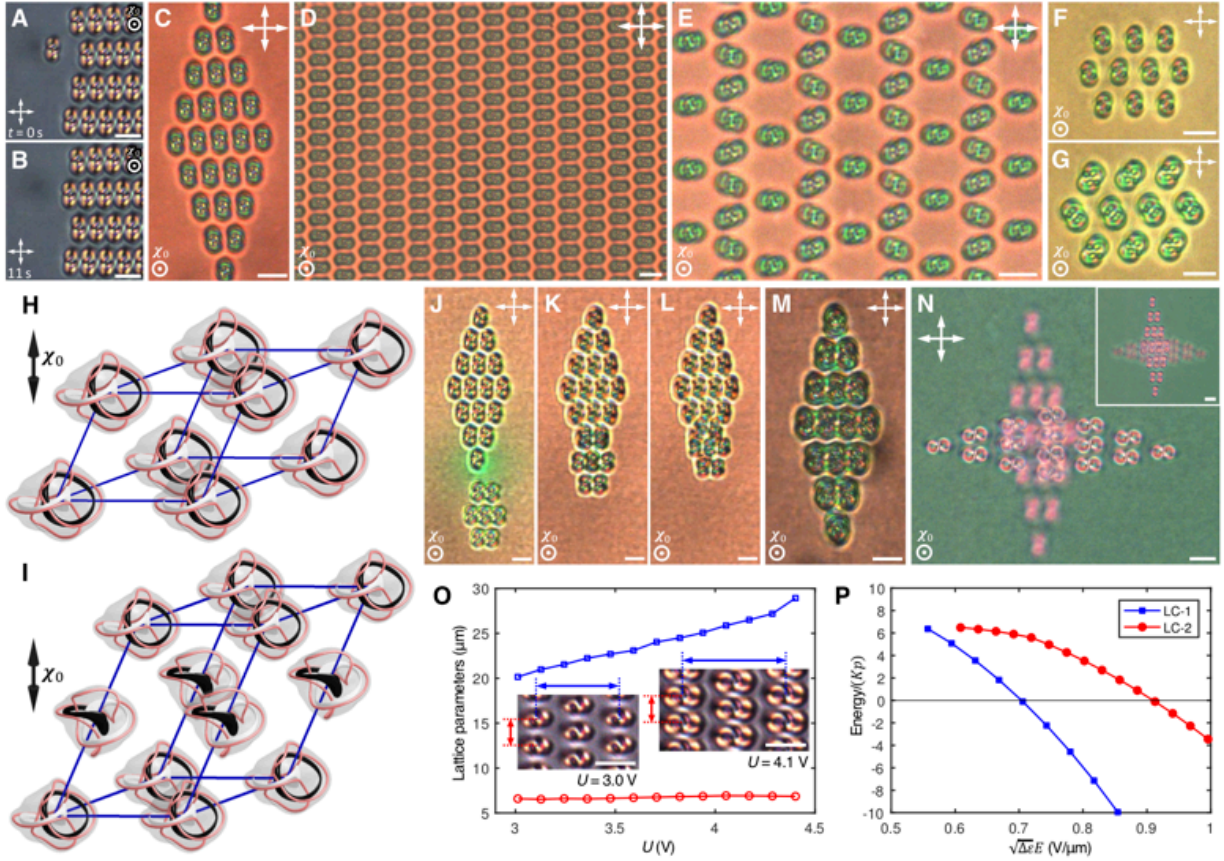

**Fig. 2. Crystals of heliknotons.** (**A** and **B**) Snapshots showing self-assembly of a 2D crystal ($d = 10\mu m$, $U$=3.5V). (**C** to **E**) 2D closed rhombic (C,D) and open (E) lattices of heliknotons at $U$=1.9V and 1.7V, respectively ($d = 15\mu m$). (**F** and **G**) Crystallites with aligned (F) and anticlinically tilted (G) heliknotons at $U$=1.8V and 2.3V, respectively ($d = 17.5\mu m$). (**H** and **I**) Primitive cells of 3D heliknoton crystals where the solitons in neighboring horizontal layers have relative parallel (H) or perpendicular (I) orientations. Isosurfaces (gray) show the localized 3D regions of heliknotons with distorted helical background when co-located with both vortex knots (light red) and preimages of anti-parallel vertical orientations in $\mathbf{n}(\mathbf{r})$ (black and white). (**J** to **L**) 3D interactions and self-assembly of heliknoton crystallites ($d \approx 30\mu m$ and $U$=2.8V). (**M** and **N**) 3D heliknoton lattices comprising crystallites with parallel (M) or perpendicular (N) orientations, where (N) and its inset are polarizing micrographs obtained when focusing at different crystalline planes ~10μm apart ($d \approx 30\mu m$ in both cases and $U$=2.8V and 3.4V, respectively). (**O**) Electrostriction of a heliknoton crystal. Insets show lattices at different $U$, with the lattice parameters shown in blue and red ($d = 10\mu m$). (**P**) Free energy of heliknoton crystals per primitive cell for two LCs at different $\mathbf{E}$. The heliknotons become metastable with respect to the unwound state at $\sqrt{\Delta\varepsilon}E \gtrsim 0.8V/\mu m$ for LC-1 and at $\sqrt{\Delta\varepsilon}E \gtrsim 1.2V/\mu m$ for LC-2. Data obtained using LC-2 in (A), (B) and (O), and LC-1 in (C-N) (*24*). Scale bars are 10μm and $p = 5\mu m$.

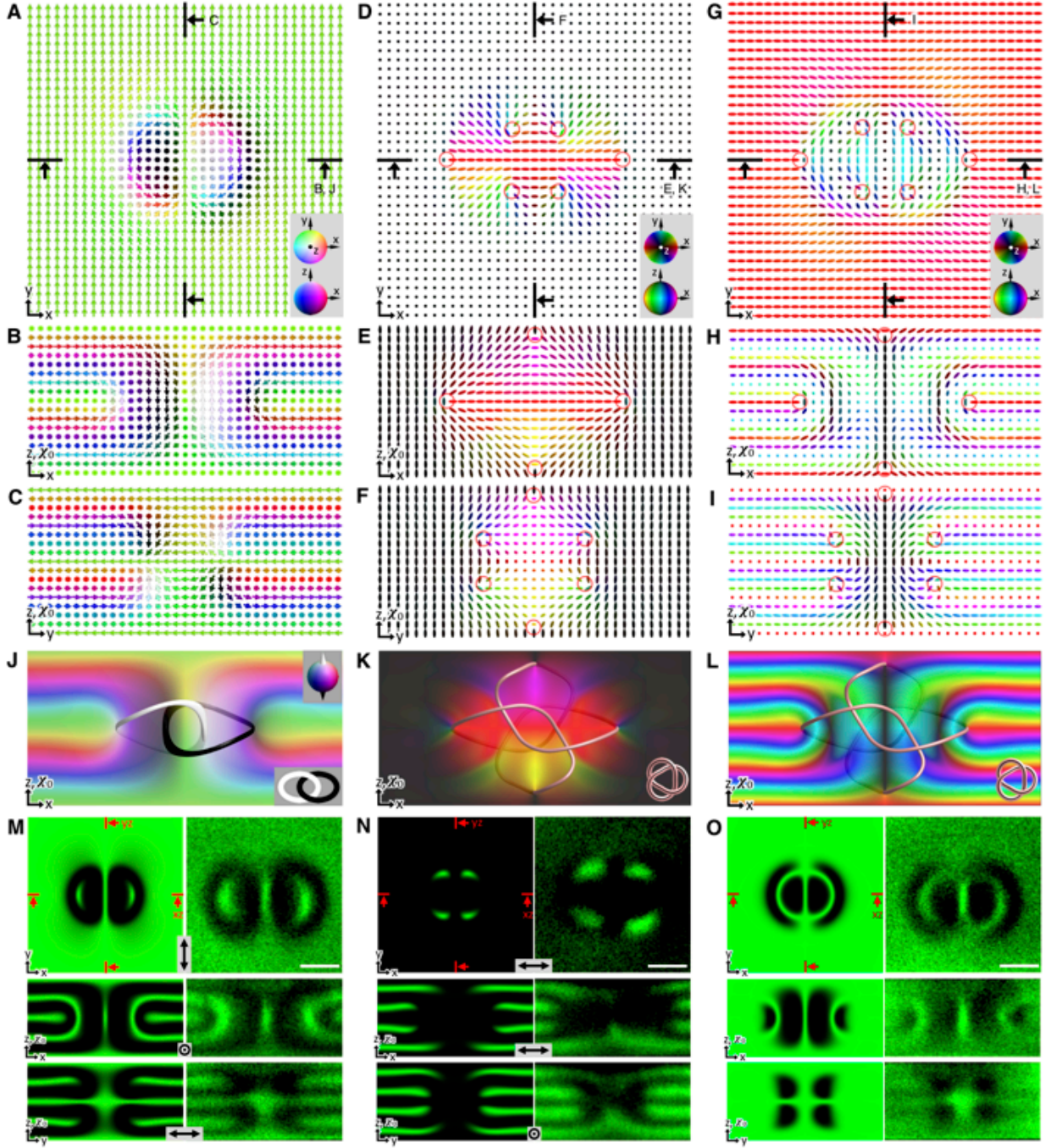

**Fig. 3. Structure of an elementary heliknoton.** (**A** to **I**) Computer-simulated cross-sections of $n(r)$ [(A) to (C)], $\chi(r)$ [(D) to (F)] and $\tau(r)$ [(G) to (I)], of a heliknoton. Vertical cross-sections and the viewing directions are marked in (A), (D), and (G), respectively. $n(r)$ is shown with arrows colored according to $\mathbb{S}^2$ [(A), Inset] and $\chi(r)$ and $\tau(r)$ are shown with ellipsoids colored according to their orientations on the doubly-colored $\mathbb{S}^2/\mathbb{Z}_2$ [(D) and (G), insets]. The vortex lines in $\chi(r)$ and $\tau(r)$ are marked by red circles in (D) to (I). (**J**) Preimages of vertical orientations of $n(r)$ forming a Hopf link (lower-right inset) and the cross-section of $n(r)$. (**K** and **L**) The singular vortex line in $\chi(r)$ and $\tau(r)$ forming a trefoil knot (lower right insets)

visualized by light-red tubes and the cross-sections of $\chi(r)$ and $\tau(r)$, respectively. (**M** to **O**) Computer-simulated and experimental nonlinear optical images of $n(r)$ in the cross-sections of a heliknoton obtained with marked linear polarizations [(M) and (N)] and circular polarization [(O)]. Left-side images are numerical and right-side ones are experimental, all obtained for LC-3 at $d = 10\mu m$ and $U=1.7V$ (*24*). Scale bars are 5 μm and $p = 5\mu m$.

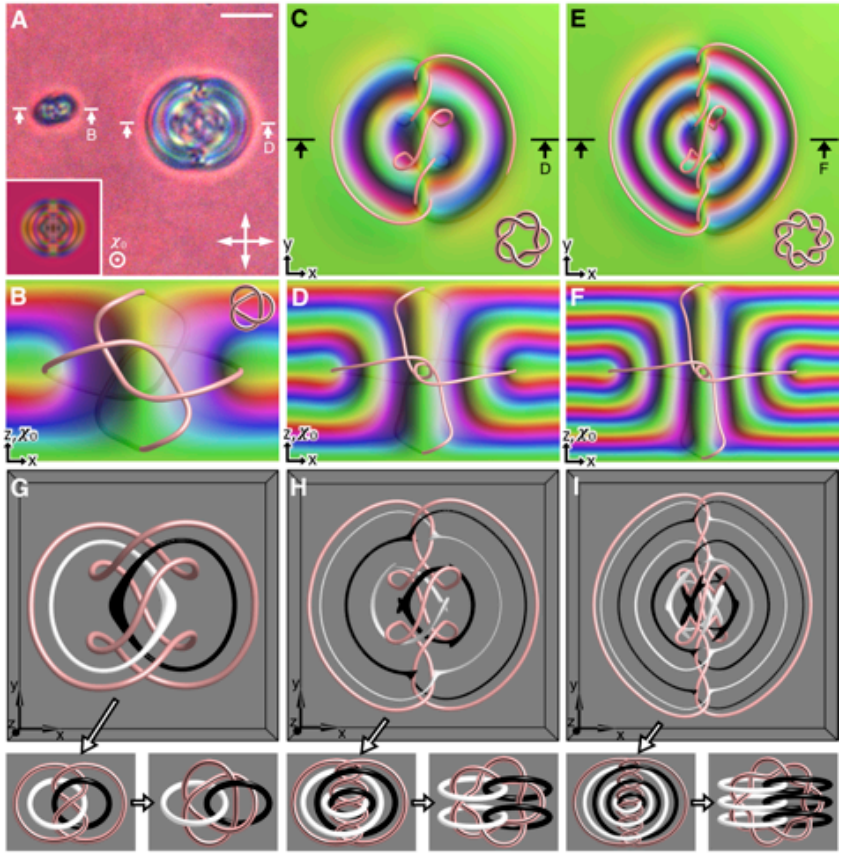

**Fig. 4. Topologically distinct heliknotons.** (**A**) Experimental polarizing optical micrograph showing an elementary $Q = 1$ (left) and a $Q = 2$ heliknoton (right), with the computer-simulated counterpart of a $Q = 2$ heliknoton shown in the inset ($d = 20\mu m$; $U=2.0V$). (**B** to **F**) Midplane cross-sections colored by $n(r)$ orientations on $\mathbb{S}^2$ and the knotted singular vortex lines in $\chi(r)$ and $\tau(r)$ shown as light-red tubes for $Q = 1$ [(B)], $Q = 2$ [(C) and (D)] and $Q = 3$ [(E) and (F)] heliknotons, respectively. The schematics of vortex lines forming right-handed T(3,2), T(5,2) and T(7,2) torus knots are shown in insets of (B), (C) and (E), respectively. (**G** to **I**) Knotted preimages of vertical orientations in $n(r)$ (black and white) and vortex lines in $\chi(r)$ and $\tau(r)$ (light red) for $Q = 1, 2$ and $3$, respectively. Shown in the lower parts of the panels are the sequences of Reidemeister moves transforming the energy-minimizing configurations into simplified links and torus knots. Scale bars are 10μm and $p = 5\mu m$; data obtained for LC-1 (*24*).

# Supplementary Materials for

## Three-dimensional crystals of adaptive knots

Jung-Shen B. Tai, Ivan I. Smalyukh[*]

Correspondence to: ivan.smalyukh@colorado.edu

**This PDF file includes:**

    Materials and Methods
    Figs. S1 to S6
    Table S1
    Captions for Movies S1 to S8

**Other Supplementary Materials for this manuscript include the following:**

    Movies S1 to S8
    An MS Excel file with the data used to obtain plots in figures

**Materials and Methods**

1. Experimental methods

1.1. Materials and sample preparation

Right-handed chiral nematic liquid crystal (LC) mixtures were prepared by mixing 4-Cyano-4'-pentylbiphenyl (5CB, from EM Chemicals) with a small amount of chiral additive, α, α, α, α-tetraaryl-1,3-dioxolane-4,5-dimethanol (TADDOL, Sigma-Aldrich) (mixture LC-1) or a low-birefringence nematic mixture ZLI-3412 (EM Chemicals) with chiral additive CB-15 (EM Chemicals) (mixture LC-2). Preparation of a 5CB-based photopolymerizable mixture (mixture LC-3) is described below in section 1.3 devoted to imaging. Left-handed chiral LCs were prepared by mixing 5CB with a left-handed chiral additive, cholesterol pelargonate (Sigma-Aldrich) and were used to confirm that heliknotons can be found in both left-and righthanded LCs. The helical pitch of the chiral LC mixture is determined by $p = (\xi c)^{-1}$, where $\xi$ is the helical twisting power of the chiral additive and $c$ is the mass fraction of the additive. In the experiments, we varied $p =$ 5-10 μm. The material parameters of 5CB, ZLI-3412 and the helical twisting power of chiral additives are listed in table S1. LC cells were assembled from indium-tin-oxide (ITO)-coated glass slides treated with polyimide PI-2555 (HD Micro-Systems) to ensure unidirectional planar alignment at the LC/glass interface, as well as the orientation of the helical axis perpendicular to substrates. Polyimide was applied to substrates by spin-coating at 2,700 revolutions per minute for 30 s and then cured by baking for 5 min at 90 °C and then 1 h at 180°C. The coated substrates were then rubbed unidirectionally to produce homogeneous planar anchoring. LC cells with uniform gap thicknesses 10-100 μm were obtained by sandwiching silica spheres in ultraviolet-curable glue between substrates. Wire leads were attached to the ITO electrodes to enable electric control. In cells with the thickness over pitch ratio $d/p \gtrsim 1.5$ (in this work studied in samples with $d/p = 1.5 - 20$) and for a range of applied voltages dependent on material parameters, heliknoton structures correspond to minima of free energy and can be generated spontaneously when filling the cells or quenching the sample from isotropic phase in presence of applied fields. They can be also obtained even more controllably using laser tweezers by locally heating the LC to isotropic phase and quenching it back to the LC state.

1.2. Diffusion, interaction of heliknotons and polarizing optical imaging

Generation and manipulation of heliknotons were done by holographic laser tweezers capable of producing arbitrary patterns of laser light intensity within the LC sample. The tweezers setup is based on an ytterbium-doped fiber laser (YLR-10-1064, IPG Photonics, operating at 1,064 nm) and a phase-only spatial light modulator (P512-1064, Boulder Nonlinear Systems) integrated with an inverted optical microscope (IX81, Olympus) (*31*). Polarizing optical microscopy and videomicroscopy were performed using the same IX-81 Olympus inverted microscope and a charge-coupled device camera (Grasshopper or Flea, both from PointGrey Research) (*31*). The polarizing optical micrographs were computer-simulated based on the Jones matrix approach (*21,31*) for free-energy-minimizing field configurations and by using the optical birefringence Δn values of the corresponding nematic hosts (table S1).

The experimental characterizations of diffusion statistics of heliknotons were done by analyzing the spatial position of a heliknoton in each video frame of a video taken at a frame rate 15 frames per second. The video frames were processed using the tracking plugins of the ImageJ software (freeware obtained from NIH) (*32*). The data were then fitted with the probability density function $P(\Delta \boldsymbol{r}, \tau) = \frac{1}{\sqrt{4\pi D \tau}} \exp\left(-\frac{(\Delta r)^2}{4D\tau}\right)$ to obtain the diffusion constant (*12,33*), where $P(\Delta \boldsymbol{r}, \tau)$ is the probability that a particle displaces by $\Delta \boldsymbol{r}$ over the elapsed time $\tau$, and $D$ is the diffusion constant.

Characterization of anisotropic pair interactions between two heliknotons was performed by releasing them from laser tweezers at a distance at which one observes little interaction, which is of strength comparable to thermal fluctuations. During attraction or repulsion, the rate of change of heliknoton separation was fitted to the highly damped equation of motion to obtain the interaction force $F_{\text{int}} = \xi dR/dt$, where $R$ is the soliton-soliton center-to-center separation distance and $\xi$ is the viscous drag coefficient derived from the aforementioned analysis of diffusion statistics through the Einstein relation, $\xi = k_B T/D$, with $k_B T$ being the product of the Boltzmann constant and temperature. The pair interaction potential was then obtained through numerical integration of the experimentally measured interaction force.

1.3. 3D nonlinear optical imaging and laser trapping

The nonlinear optical imaging of the material alignment field $\boldsymbol{n}(\boldsymbol{r})$ within heliknotons was performed by using a three-photon excitation fluorescence polarizing microscopy (3PEF-PM) set-up built around the same multimodal IX-81 Olympus inverted microscope as that integrated with the laser tweezers described above (*19*). The integration of nonlinear optical imaging modality, polarizing optical microscopy modality, and laser tweezers in the same multimodal optical setup allows for in-situ optical generation, non-contact control, and nondestructive 3D imaging of the heliknotons. To reduce imaging artifacts due to the birefringence of LCs and maintain the integrity of the soliton structure during the imaging process, a partially polymerizable chiral LC mixture (mixture LC-3) was prepared by mixing 84% of 5CB with 15% of diacrylate nematic RM 257 (Merck) and 1% of UV-sensitive photoinitiator Irgacure 369 (Sigma-Aldrich). Before imaging, the LC sample containing heliknotons was photo-polymerized under weak UV illumination to avoid perturbations from thermal gradient or optical realignment of LC molecules, and the unpolymerized 5CB molecules were washed away and replaced by immersion oil. In some cases, a trace amount of RM257 was added to the spin-coating process of PI-2555 on the substrates to aid in the adherence of the polymerized LC film to the substrates. This process reduced the effective birefringence by an order of magnitude, thus minimizing imaging artifacts such as beam defocusing and polarization changes (*34*). The remaining 5CB molecules were excited via three-photon absorption by using a Ti-Sapphire oscillator (Chameleon Ultra II; Coherent) operating at 900 nm with 140-fs pulses at a repetition rate of 80 MHz (*19*). The fluorescence signal was epi-detected by using a 417/60-nm bandpass filter and a photomultiplier tube (H5784-20, purchased from Hamamatsu). An oil-immersion 100X objective with NA = 1.4 was used. The polarization state of the excitation beam was controlled by using a polarizer and a rotatable half-wave retardation plate or a quarter-wave retardation plate. No polarizers were utilized at the detection channel. The 3PEF-PM involves a third-order nonlinear process and its intensity scales as $\cos^6 \beta$, where $\beta$ is the angle between the dipole moment of the LC molecule, orienting along $\boldsymbol{n}(\boldsymbol{r})$, and the polarization direction of the excitation beam (*19,21*). The 3D 3PEF-PM images for different polarizations of the excitation light were obtained by scanning the excitation beam through the sample volume and recording the fluorescence intensity as a function of 3D scanning coordinates. The images were then post-processed by background subtraction, depth-dependent intensity normalization, and contrast enhancement. For a given linear polarization of the beam, each 3PEF-PM image yielded a preimage of a single point on $\mathbb{S}^2/\mathbb{Z}^2$ or preimages of a pair of diametrically opposite points on $\mathbb{S}^2$ for vector $\mathbf{n}(\mathbf{r})$ field, due to the nonpolar response of molecules in LC to the excitation (*23,35-37*). For a vectorized director field (i.e., a director field smoothly decorated by vectors), the preimages can be assigned by comparing the experimental preimages to preimages obtained in numerical modeling. Computer simulations of the 3PEF-PM images were based on the $\propto \cos^6 \beta$ dependence of the fluorescence image intensity on the molecular director orientation relative to the polarization direction of the excitation light. The direct comparison (Fig. 3) of experimental and computer-simulated images was used to unambiguously confirm that the experimentally reconstructed heliknoton structures are indeed the ones corresponding to their numerical counterparts.

1.4. Sample preparation for self-assembly of 2D and 3D crystals of heliknotons

The anisotropic attractive interaction between heliknotons allow the formation of various two-dimensional crystals by self-assembly and/or laser-tweezer-guided assembly. In cells with $1.5 \lesssim d/p \lesssim 4$, heliknotons are constrained by short-range elastic repulsions from confining substrates due to the boundary conditions. Therefore, the heliknotons equilibrate around the horizontal midplane of the cell, with the interaction between heliknotons being quasi-2D. At $d/p \gtrsim 4$, heliknotons also diffuse and move along $\chi_0$ in response to thermal fluctuations and when manipulated with laser tweezers, accompanied by the helical rotation (Fig. 1, D and E, and movie S2). The 3D crystalline assemblies were studied in cells with $d/p = 4 - 20$.

2. Methods to characterize heliknoton stability and topology

2.1. Free energy minimization

The energetic cost of spatial deformations of $\mathbf{n}(\mathbf{r})$ in a chiral nematic LC can be described by the Frank-Oseen free energy functional (*12,21*)

$$F_{\text{elastic}} = \int d^3 r \left\{ \frac{K_{11}}{2} (\nabla \cdot \mathbf{n})^2 + \frac{K_{22}}{2} [\mathbf{n} \cdot (\nabla \times \mathbf{n})]^2 + \frac{K_{33}}{2} [\mathbf{n} \times (\nabla \times \mathbf{n})]^2 + \frac{2\pi K_{22}}{p} \mathbf{n} \cdot (\nabla \times \mathbf{n}) \right.$$
$$\left. - \frac{K_{24}}{2} \{\nabla \cdot [\mathbf{n}(\nabla \cdot \mathbf{n}) + \mathbf{n} \times (\nabla \times \mathbf{n})]\} \right\}, \quad (S1)$$

where the Frank elastic constants $K_{11}$, $K_{22}$, $K_{33}$, and $K_{24}$ describe the energetic costs of splay, twist, bend and saddle-splay deformations, respectively. When an external electric field is applied, an additional dielectric coupling term in the free energy has to be included due to the dielectric properties of the LCs, so that the elastic energy in Eq. (S1) is supplemented with the corresponding electric field coupling term,

$$F_{\text{electric}} = -\frac{\varepsilon_0 \Delta \varepsilon}{2} \int d^3 r (\mathbf{E} \cdot \mathbf{n})^2, \quad (S2)$$

where $\varepsilon_0$ is the vacuum permittivity, $\Delta \varepsilon$ is the dielectric anisotropy of the LC and $\mathbf{E}$ is the applied electric field. The total free energy, which is the sum of elastic and electric coupling energies given by a sum of equations (S1) and (S2) can be reduced to a simpler form of Eq. (1) in the main text by adopting the one-elastic-constant approximation; this is usually a good approximation because of the anisotropy in elastic constants being small for LCs (table S1). Note that the first two terms in Eq. (1), though with different definitions of constants/variables, also describe the micromagnetic

Hamiltonian of chiral ferromagnets (*14, 22, 23, 37*). In this work, we perform numerical simulations for the material parameters given in table S1, which includes the elastic constant anisotropy. However, we also find that the heliknotons can be stabilized within the one-elastic-constant approximation, which means that they can potentially exist in a broad range of materials, including the non-centrosymmetric magnets.

For a localized field configuration $\mathbf{n}(\mathbf{r})$ to exist, it needs to emerge as a local or global minimum of the free energy given by the sum of Eqs. (S1) and (S2). Numerical modeling of the energy-minimizing $\mathbf{n}(\mathbf{r})$ structures is performed using a variational-method-based relaxation routine implemented using the common finite differences approach (*21-23, 31*). For example, at each iteration of the numerical simulation, $\mathbf{n}(\mathbf{r})$ is updated based on an update formula derived from the Euler-Lagrange equation of the system, $n_i^{\text{new}} = n_i^{\text{old}} - \frac{\text{MSTS}}{2}[F]_{n_i}$, where the subscript $i$ denotes spatial coordinates, $[F]_{n_i}$ denotes the functional derivative of $F$ with respect to $n_i$, and MSTS is the maximum stable time step in the minimization routine, determined by the values of elastic constants and the spacing of the computational grid (*31*). The steady-state stopping condition is determined by monitoring the change in the spatially averaged functional derivatives over iterations. When this value approaches zero, it implies the system is in a state corresponding to an energy minimum and the relaxation routine is terminated.

The Frank elastic constants adopted in the numerical modeling are based on values for the two nematic hosts used in this study (table S1) (*19,21,23*). The simulations with the saddle-splay elastic constant $K_{24} = K_{22}$ and $K_{24} = 0$ yielded similar results, consistent with the absence of singular defects (*21,31*). The 3D spatial discretization is performed on dense 3D square grids, with one helicoidal pitch $p$ distance equivalent sampled by 24 or 32 grid points. The spatial derivatives are calculated using the finite difference methods with second-order accuracies, allowing us to minimize discretization-related artifacts in modeling of the structures of the solitons. The cell thickness values $d$ in simulations were between 10 to 100 μm and $p = 5$ μm. Periodic boundary conditions are set on lateral boundaries of the simulation volume, and unidirectional or periodic boundary conditions are set on the top and bottom boundaries. When simulating individual heliknotons, the lateral dimension of the simulation volume is set to be sufficiently large for the field configuration $\mathbf{n}(\mathbf{r})$ on the periphery to fully relax to the helical background. Initial conditions for energy-minimizing relaxations of heliknotons were inspired by experimental optical imaging and the stabilized field configurations were consistent with the experimental results.

To construct a preimage of a point on $\mathbb{S}^2$ within the 3D volume of the static topological solitons, we calculate a scalar field defined as the difference between the solitonic field $\mathbf{n}(\mathbf{r})$ and a unit vector defined by the target point on $\mathbb{S}^2$. The preimage is then visualized by the isosurfaces of a small value in this ensuing scalar field (*21,23*). The freely available software KnotPlot (*38*) is used for simplifying and then visualizing linking of preimages.

2.2. Construction of $\boldsymbol{\chi}(\mathbf{r})$ and $\boldsymbol{\tau}(\mathbf{r})$ based on the material field $\mathbf{n}(\mathbf{r})$

The nonpolar line field of helical axis $\boldsymbol{\chi}(\mathbf{r})$ was derived from the material director field $\mathbf{n}(\mathbf{r})$ by identifying the twist axis at each spatial position in the simulation volume based on the left eigenvector of the chirality tensor

$$\chi_i C_{ij} = \lambda \chi_j, \quad (S3)$$

where $C_{ij} = n_k \epsilon_{ljk} \partial_i n_l$ is the chirality tensor and $\lambda$ is the local eigenvalue (*29,30*). The helical axis $\boldsymbol{\chi}(\mathbf{r})$ thus calculated indicates the direction of twisting and is nonpolar in nature. Singularities in $\chi(r)$ field were determined by finding regions where the twist axis is ill-defined. Singular vortex lines were visualized by a tube following the singular vortex in $\chi(r)$. The third field $\tau(r)$ is perpendicular to both $n(r)$ and $\chi(r)$, and was constructed to be the nonpolar line field along the cross product $n(r) \times \chi(r)$.

2.3. Characterization of topology of the heliknotons

In the material field $n(r)$, heliknotons are localized 3D configurations of smooth fields embedded in a uniform helical background with a helical axis $\chi_0$. The uniform helical background allows the compactification of the configuration space (domain space) $\mathbb{R}^3$ to a 3-sphere $\mathbb{S}^3$. If considering only the material field $n(r)$, the target space is the order-parameter space of directors $n(r)$, namely a sphere with antipodal points identified, $\mathbb{S}^2/\mathbb{Z}_2$, due to the head-tail symmetry of LC alignment field. The field configuration is then characterized by topologically distinct $\mathbb{S}^3 \to \mathbb{S}^2/\mathbb{Z}_2$ maps; they belong to the third homotopy group $\pi_3(\mathbb{S}^2/\mathbb{Z}_2) = \mathbb{Z}$ (*3,21,23,39*). Since continuous line fields in a simply connected space can always be smoothly decorated with arrows into unit vector fields and $\pi_3(\mathbb{S}^2/\mathbb{Z}_2)$ and $\pi_3(\mathbb{S}^2)$ are isomorphic (*16,40*), the material field $n(r)$ of a heliknoton can be treated equivalently as a unit vector field, with heliknotons in this vectorized field labeled as the elements of $\pi_3(\mathbb{S}^2) = \mathbb{Z}$. The action of smoothly vectorizing a line field in LC materials physically corresponds to dispersing polar nanoparticles, such as ferromagnetic nanoplates, to achieve ferromagnetic order in the ensuing composite material (*23,37*). The homotopy groups $\pi_3(\mathbb{S}^2/\mathbb{Z}_2) = \mathbb{Z}$ (or $\pi_3(\mathbb{S}^2) = \mathbb{Z}$ ) suggest topologically distinct field configurations in a helical background can be characterized by the Hopf index $Q \in \mathbb{Z}$. For a unit vector field, the Hopf index bears a geometric interpretation as the linking number of any two closed-loop preimages, regions in space with the same orientation of field corresponding to a single point on $\mathbb{S}^2$ (*40*). As an example, this is shown in Fig. 3J and fig. S1 where preimages of any two distinct points on $\mathbb{S}^2$ are linked exactly once for an elementary heliknoton. Interesting examples in this case are the preimages of points on the equator of $\mathbb{S}^2$, which correspond to the helical far-field background. Hopf indices of a heliknoton can also be evaluated numerically by integrating in the configuration space $\mathbb{R}^3$ according to Eq. (2) in the main text (*40-43*). By defining $b^i \equiv \epsilon^{ijk}B_{jk}$, one gets $b^i = \frac{1}{2}\epsilon^{ijk}(\partial_j A_k - \partial_k A_j) = \epsilon^{ijk}\partial_j A_k$ and $A$ can be understood as the vector potential of the vector field $b$, since $b = \nabla \times A$, and $Q$ can be rewritten as $Q = \frac{1}{64\pi^2}\int d^3r\, b \cdot A$. After calculating $b$ from $n(r)$, the vector potential $A$ can be obtained by the direct integration of $b$ (*22,37*). Hopf index values obtained by means of geometric analysis of preimage linking and numerical integration closely match, up to numerical precision.

When considering all three orthonormal helical fields, $\tau(r) \perp n(r) \perp \chi(r)$, the target space becomes $\mathbb{S}^3/\mathbb{Z}_2$ [SO(3)] when all three fields are polar. The order-parameter space is characterized by $\mathbb{S}^3/Q_8$ [SO(3)/$D_2$] when all three fields are nonpolar, where $Q_8$ and $D_2$ are the quaternion group and the dihedral group of order 4, respectively (*44,45*). In addition to the chiral nematics, orthorhombic biaxial nematics also have a similar target space $\mathbb{S}^3/Q_8$. We note that the material field configurations of heliknotons we study are nonsingular as a result of the two out of three fields being immaterial and nonpolar, as in the case of chiral nematics. For biaxial nematics, all three fields are material fields. Energetically, singular vortex knots in material fields would be hard to stabilize without specific boundary conditions or colloidal inclusions, though LC chirality potentially could also enable such stabilization.

2.4. Analysis of unwinding and undulation instabilities at high fields

A simple comparison of free energies reveals that the bulk chiral LC samples with helical structures can be unwound at high applied electric fields, which correspond to $\sqrt{\Delta\varepsilon}E \gtrsim 0.73$V/μm for LC-1 and at $\sqrt{\Delta\varepsilon}E \gtrsim 1.08$V/μm for LC-2. Because of this, the individual heliknotons at high voltages would be effectively embedded in the uniform director background and correspond to hopfions *(3,4,6)*. The 3D heliknoton crystals are stable with respect to unwinding up to $\sqrt{\Delta\varepsilon}E \gtrsim 0.8$V/μm for LC-1 and at $\sqrt{\Delta\varepsilon}E \gtrsim 1.2$V/μm for LC-2, becoming metastable and eventually unstable at higher fields. In LC cells of finite thickness with helical axis aligned perpendicular to confining substrates with strong surface anchoring, the stability range is increased further due to the effects of confinement. For example, for LC-1, the heliknotons in cells of thickness of 10μm are stable up to $\sqrt{\Delta\varepsilon}E = 1.04$V/μm and metastable at even higher fields. In addition to unwinding, the cholesteric helical structures can undergo undulation instabilities *(46)*, which arise from the competition of the aligning action of electric field (tending to orient $\boldsymbol{n}(\boldsymbol{r})\|\boldsymbol{E}$) and confining surfaces that tend to keep the helical axis perpendicular to substrates and along the applied $\boldsymbol{E}$. These undulating cholesteric structures are topologically equivalent to the undistorted helical state and can host heliknotons within them. These undulation patterns can potentially provide further means for controlling spatial organizations of heliknotons.

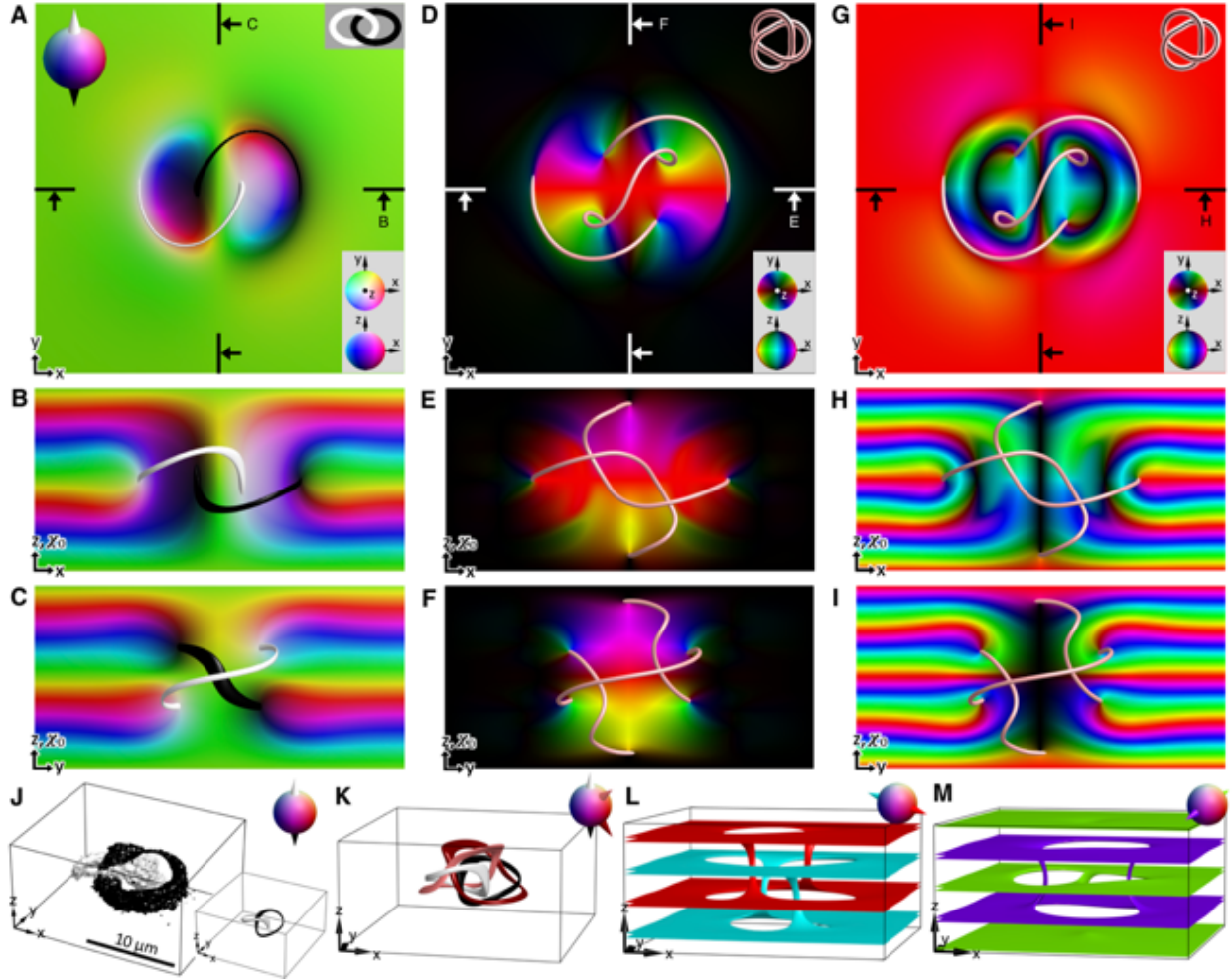

**Fig. S1. Details of the elementary heliknoton's structure.** (**A** to **I**) Computer-simulated midplane cross-sections of the triad fields, $\boldsymbol{n}(\boldsymbol{r})$ [(A) to (C)], $\boldsymbol{\chi}(\boldsymbol{r})$ [(D) to (F)] and $\boldsymbol{\tau}(\boldsymbol{r})$ [(G) to (I)], of an elementary heliknoton shown in Fig. 3. Vertical cross-sections and the viewing directions are marked in (A), (D) and (G), respectively. $\boldsymbol{n}(\boldsymbol{r})$ is shown with colors according to their orientations as points on $\mathbb{S}^2$ [(A), Inset] and line fields $\boldsymbol{\chi}(\boldsymbol{r})$ and $\boldsymbol{\tau}(\boldsymbol{r})$ are shown with colors according to their orientations on the doubly-colored $\mathbb{S}^2/\mathbb{Z}_2$ [(D) and (G), insets]. In (A) to (C), preimages of anti-parallel vertical orientations are superposed with the cross-sections. In (D) to (I), singular vortex lines shown as light-red tubes are superposed with the cross-sections. (**J** to **K**) Interlinked preimages of $\boldsymbol{n}(\boldsymbol{r})$ orientations shown as cones on $\mathbb{S}^2$ in their insets. In (J), experimentally constructed preimages are shown with the simulated counterparts (lower right). Numerical and experimental data were obtained with parameters for 5CB-based LC-1 (table S1) with right-handed chirality and $p = 5$ μm at $U = 1.9$ V.

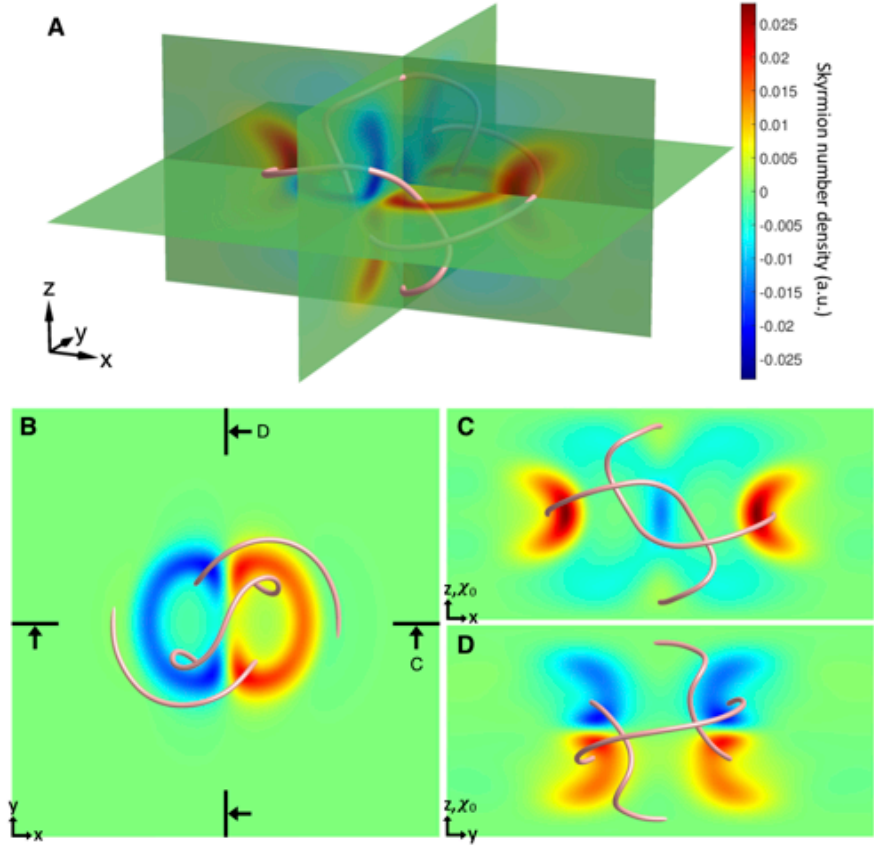

**Fig. S2. Skyrmion number density and the meron knot within a heliknoton.** (**A**) 3D view of the singular vortex knot (light-red tube) and the skyrmion number density in midplane cross-sections. (**B** to **D**) Skyrmion number density in three midplane cross-sections. The vertical cross-sections [(C) and (D)] and the viewing directions are marked in (B). The Skyrmion number in a 2D plane with a surface normal $\hat{\alpha}$ is $N_\alpha = \frac{1}{8\pi}\int d^2r\, \epsilon^{\alpha\beta\gamma} \boldsymbol{n}(\boldsymbol{r}) \cdot (\partial_\beta \boldsymbol{n}(\boldsymbol{r}) \times \partial_\gamma \boldsymbol{n}(\boldsymbol{r}))$ (*47*) where $\alpha$, $\beta$ and $\gamma$ are spatial coordinates and $\epsilon$ is the totally antisymmetric tensor. Surface normals are $+\hat{\boldsymbol{x}}$, $+\hat{\boldsymbol{y}}$ and $+\hat{\boldsymbol{z}}$ for the three midplane cross-sections. Each lump in the cross-sections around the vortex line corresponds to one half of skyrmion number, forming a meron. Numerical data were obtained with parameters for 5CB-based LC-1 with right-handed chirality and $p = 5$ µm at $U = 1.9$ V.

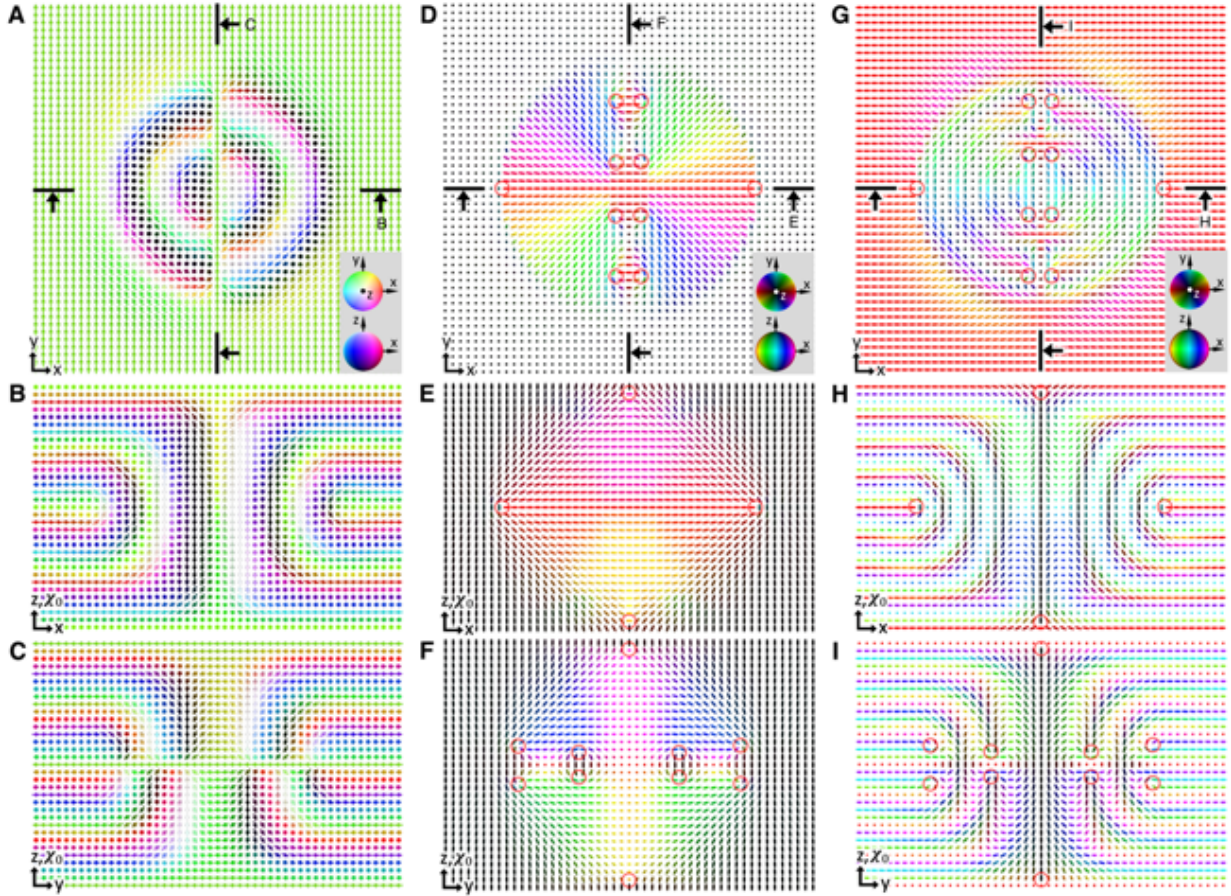

**Fig. S3. Details of the $Q = 2$ heliknoton structure.** (A to I) Computer-simulated midplane cross-sections of the triad fields, $n(r)$ [(A) to (C)], $\chi(r)$ [(D) to (F)] and $\tau(r)$ [(G) to (I)], of a $Q = 2$ heliknoton shown in Fig. 4, A, C and D. The vertical cross-sections and the viewing directions are marked in (A), (D) and (G), respectively. $n(r)$ is shown with arrows colored according to their orientations on $\mathbb{S}^2$ [(A), Inset] and $\chi(r)$ and $\tau(r)$ are shown with ellipsoids colored according to their orientations on the doubly-colored $\mathbb{S}^2/\mathbb{Z}_2$ [(D) and (G), insets]. The vortex lines in $\chi(r)$ and $\tau(r)$ intersecting the cross-sectional planes are marked by red circles in (D) to (I). Numerical data were obtained with parameters for 5CB-based chiral LC-1 with right-handed chirality and $p = 5$ μm at $U= 2.5$ V.

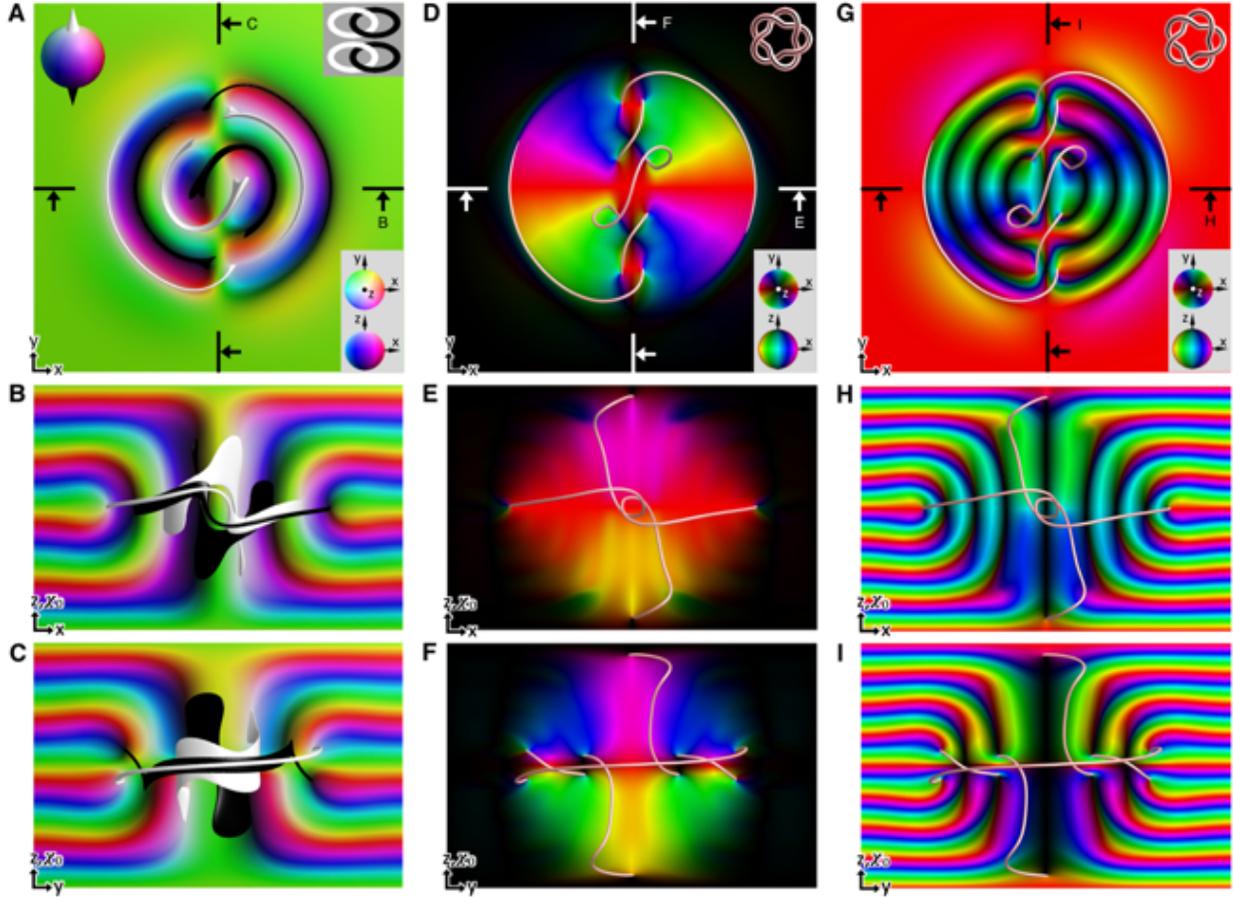

**Fig. S4. Cross-sections, preimages, and singular vortex lines of the $Q = 2$ heliknoton.** (A to I) Computer-simulated midplane cross-sections of the triad fields, $n(r)$ [(A) to (C)], $\chi(r)$ [(D) to (F)] and $\tau(r)$ [(G) to (I)], of a $Q = 2$ heliknoton in Fig. 4, A, C and D. The vertical cross-sections and the viewing directions are marked in (A), (D) and (G), respectively. $n(r)$ is shown with colors according to their orientations on $\mathbb{S}^2$ [(A), Inset] and $\chi(r)$ and $\tau(r)$ are shown with colors according to their orientations on the $\mathbb{S}^2/\mathbb{Z}_2$ [(D) and (G), insets]. In (A) to (C), preimages of anti-parallel vertical orientations are superposed with the cross-sections. In (D) to (I), singular vortex lines shown as light-red tubes are superposed with the cross-sections. Numerical data were obtained with parameters for the 5CB-based LC-1 with right-handed chirality and $p = 5$ μm at $U = 2.5$ V.

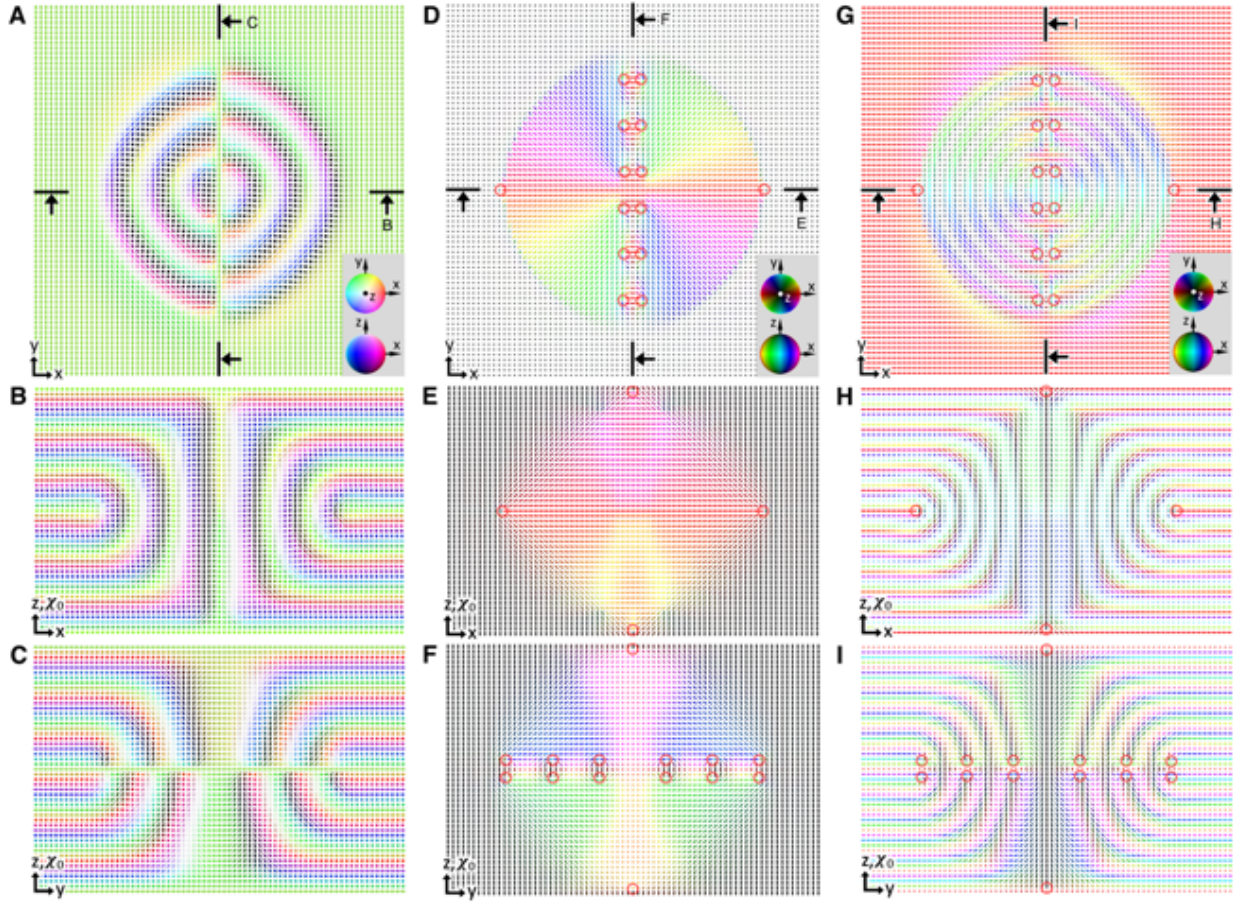

**Fig. S5. Details of the $Q = 3$ heliknoton's structure.** (**A** to **I**) Computer-simulated midplane cross-sections of the triad fields, $n(r)$ [(A) to (C)], $\chi(r)$ [(D) to (F)] and $\tau(r)$ [(G) to (I)], of a $Q = 3$ heliknoton shown in Fig. 4, E and F. The vertical cross-sections and the viewing directions are marked in (A), (D) and (G), respectively. $n(r)$ field is shown with arrows colored according to their orientations on $\mathbb{S}^2$ [(A), Inset] and $\chi(r)$ and $\tau(r)$ are shown with ellipsoids colored according to their orientations as points on the doubly-colored $\mathbb{S}^2/\mathbb{Z}_2$ [(D) and (G), insets]. The vortex lines in $\chi(r)$ and $\tau(r)$ are marked by red circles in (D) to (I). Numerical data were obtained with parameters of the 5CB-based chiral LC-1 with right-handed chirality and $p = 5$ μm at $U = 2.8$ V.

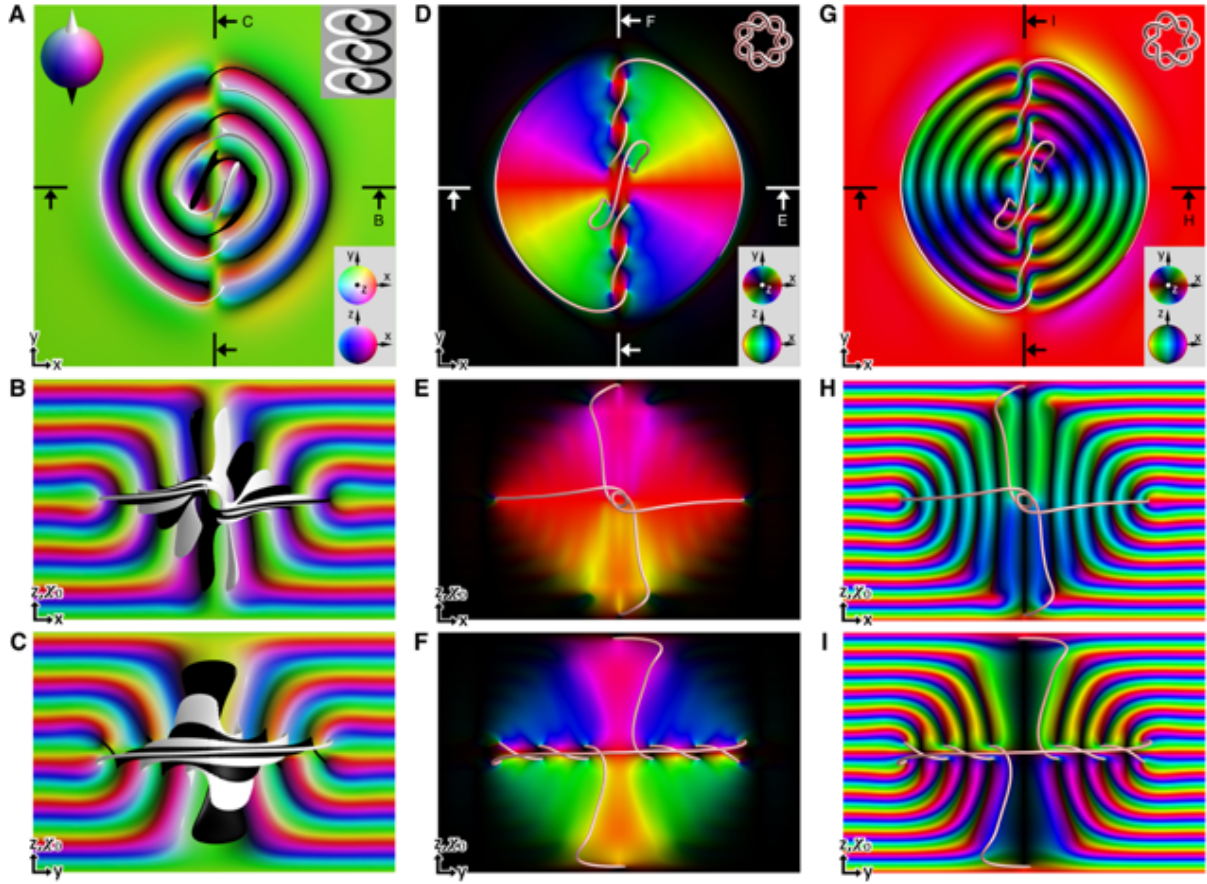

**Fig. S6. Cross-sections, preimages, and singular vortex lines within the $Q = 3$ heliknoton.**
(**A** to **I**) Computer-simulated midplane cross-sections of the triad fields, $n(r)$ [(A) to (C)], $\chi(r)$ [(D) to (F)] and $\tau(r)$ [(G) to (I)], of a $Q = 3$ heliknoton shown in Fig. 4, E and F. The antiparallel vertical cross-sections and the viewing directions are marked in (A), (D) and (G), respectively. $n(r)$ is shown with colors according to their orientations on $\mathbb{S}^2$ [(A), Inset] and $\chi(r)$ and $\tau(r)$ are shown with colors according to their orientations on the doubly-colored $\mathbb{S}^2/\mathbb{Z}_2$ [(D) and (G), insets]. In (A) to (C), preimages of vertical orientations are superposed with the cross-sections. Numerical data were obtained with parameters for 5CB-based chiral LC-1 with right-handed chirality and $p = 5$ μm at $U = 2.8$ V.

**Table S1.**

| Nematic host material | $K_{11}$ (pN) | $K_{22}$ (pN) | $K_{33}$ (pN) | $K_{24}$ (pN) | $K$ (pN) | $\Delta\varepsilon$ | $\Delta n$ | $\xi$ of CB-15 ($\mu m^{-1}$) | $\xi$ of TADDOL ($\mu m^{-1}$) | $\xi$ of cholesterol pelargonate ($\mu m^{-1}$) |
|---|---|---|---|---|---|---|---|---|---|---|
| 5CB | 6.4 | 3 | 10 | 0 | 6.5 | 13.8 | 0.18 | 7.3 | 200 | -6.25 |
| ZLI-3412 | 14.1 | 6.7 | 15.5 | 0 | 12.1 | 3.4 | 0.08 | 6.3 | — | — |

Material parameters of nematic LCs and the helical twisting power of chiral additives used.

**Movie S1.**

Brownian motion of individual heliknotons:
The video shows Brownian motion of individual heliknotons in LC-1 (left) and LC-2 (right), respectively, observed in a microscope between crossed polarizers. The video was obtained for LC-1 at $U = 1.2$ V (left) and LC-2 at $U = 4.3$ V (right); $p = 5$ μm. The elapsed time, the orientation of crossed polarizers and $\chi_0$ are marked on the video frames.

**Movie S2.**

Three-dimensional pair interaction of two LC heliknotons:
The video shows two heliknotons interacting and assembling into a dimer by displacement in lateral directions and orientation-correlated displacement along $\chi_0$. The focus was varied by ~10 μm after 26 s in the video to show the difference in depth along $\chi_0$ of two heliknotons. The video was obtained for LC-2; $U = 11.3$ V; $p = 5$ μm. The elapsed time and the orientations of crossed polarizers and $\chi_0$ are marked on the video frames.

**Movie S3.**

Self-assembly of a 2D heliknoton crystal:
A video showing the self-assembly process of a 2D rhombic crystal of heliknotons. The video was obtained for LC-2; $U = 3.5$ V; $p = 5$ μm. The elapsed time and the orientation of cross polarizers and $\chi_0$ are marked on the video frames.

**Movie S4.**

Electric reconfiguration of a heliknoton crystallite from synclinic to anticlinic tilting:
The video shows a 2D heliknoton crystal switched from unidirectional alignment of heliknotons at $U = 1.8$ V to the anticlinic tilting state by increasing the applied voltage to $U = 2.4$ V and then lowering it again to $U = 2.3$ V; $p = 5$ μm. The video was obtained for LC-1. The elapsed time and orientations of crossed polarizers and $\chi_0$ are marked on the video frames.

**Movie S5.**

3D interaction and self-assembly of heliknoton crystallites:
The video shows the process of 3D self-assembly of heliknoton crystallites with parallel relative orientations, guided by laser tweezers. The video was obtained for LC-1; $U = 2.8$ V. The elapsed time and orientations of crossed polarizers and $\chi_0$ are marked on the video frames. $p = 5$ μm.

**Movie S6.**

3D crystallites of heliknotons:
The video shows a 3D structure of heliknotons assembled from two 2D crystallites having perpendicular relative orientations. The two layers are displaced along $\chi_0$ by ~10 μm. In the video, the focus steps through the depth of the sample to show the spatially displaced crystallites of heliknotons. The video was obtained for LC-1; $U = 3.4$ V. The depth of the focal plane relative to the top heliknoton crystallite and the orientations of crossed polarizers and $\chi_0$ are marked on the video frames. $p = 5$ μm.

**Movie S7.**

Electrostriction of a heliknoton crystal:
The video shows a crystal of heliknotons undergoing electrostriction when $U$ is increased from 1.8 V to 2.3 V. The video was obtained for LC-1; orientations of crossed polarizers and $\chi_0$ are marked on the video frames. $p = 5$ μm.

**Movie S8.**

Electrostriction of a large heliknoton crystal:
The video shows a large crystal of heliknotons undergoing electrostriction when $U$ is increased slowly from 1.6 V to 2.1 V. The video was obtained for LC-1; orientations of crossed polarizers and $\chi_0$ are marked on the video frames. $p = 5$ μm.